# Electronic and Optical Properties of the Recently Synthesized 2D Vivianites (Vivianenes): Insights from First-Principles Calculations

Raphael Benjamim de Oliveira[a], Bruno Ipaves[a], Guilherme da Silva Lopes Fabris[a], Surbhi Slathia[b], Marcelo Lopes Pereira Júnior[c], Raphael Matozo Tromer[d], Chandra Sekhar Tiwary[b,e], Douglas Soares Galvão[a]

[a] Applied Physics Department and Center for Computational Engineering & Sciences, State University of Campinas, Campinas, São Paulo 13083970, Brazil.

[b] School of Nanoscience and Technology, Indian Institute of Technology Kharagpur, Kharagpur 721302, India.

[c] University of Brasília, College of Technology, Department of Electrical Engineering, Brasilia 70910900, Federal District, Brazil.

[d] University of Brasília, Institute of Physics, Brasilia 70910900, Federal District, Brazil.

[e] Department of Metallurgical and Materials Engineering, Indian Institute of Technology Kharagpur, Kharagpur, West Bengal 721302, India.

**Abstract:** Vivianite ($Fe_3(PO_4)_2 \cdot 8H_2O$) is a naturally occurring layered material with significant environmental and technological relevance. This work presents a comprehensive theoretical investigation of its two-dimensional (2D) counterpart, Vivianene, focusing on its structural, electronic, and optical properties. Using density functional theory (DFT) calculations and *ab initio* molecular dynamics (AIMD) simulations, we evaluate its thermodynamic stability, band structure, density of states, and optical response. Our results confirm that Vivianene retains the main structural features of bulk Vivianite while exhibiting enhanced thermodynamic stability at room temperature. The electronic structure analysis reveals an indirect bandgap of 3.03 eV for Vivianene, which is slightly lower than the 3.21 eV observed for bulk Vivianite, deviating from the expected quantum confinement trend in 2D materials. The projected density of states (PDOS) analysis indicates that Fe *d* orbitals predominantly contribute to the valence and conduction bands. Optical calculations demonstrate that Vivianene exhibits a higher optical band gap (3.6 eV) than bulk Vivianite (3.2 eV), with significant absorption in the ultraviolet region. The refractive index and reflectivity analyses suggest that most of the incident light is absorbed rather than reflected, reinforcing its potential for optoelectronic applications. These findings provide valuable insights into the fundamental properties of Vivianene and highlight its potential for advanced applications in sensing, optoelectronics, and energy-related technologies.

# Introduction

Graphene has created a revolution in materials science since its synthesis in 2004. Since then, extensive investigations have been carried out into new two-dimensional (2D) materials, revealing a range of emergent properties when materials are reduced to a single or a few atomic layers. These properties, including a high surface-area-to-volume ratio, tunable electronic structures, and enhanced mechanical flexibility, make 2D materials promising candidates for various technological applications, from sensors and energy storage to optoelectronic devices [1-3]. Given the growing interest in expanding the library of 2D materials beyond graphene, significant attention has been directed toward layered materials that can be exfoliated into their monolayer forms [1-2].

Due to their abundance and environmental significance, vivianites ($Fe_3(PO_4)_2 \cdot 8H_2O$) are an important family among naturally occurring layered materials. These ferrous phosphate minerals are commonly found in anoxic environments, where they play a fundamental role in the phosphorus cycle and contaminant immobilization [4,5]. Similar to other layered materials, such as hematene (2D $\alpha$-$Fe_2O_3$) [6], ilmenene (2D $FeTiO_3$) [7], and chromiteen (2D $FeCr_2O_4$) [8], vivianites possess a crystalline structure that allows for exfoliation into its 2D form, generically referred to as vivianenes. The exfoliation process exploits weak interlayer interactions, facilitating perfect basal cleavage and isolating stable monolayers with potentially novel properties [1,5].

The electronic properties of vivianites have been previously studied using density functional theory (DFT) calculations, which have estimated their electronic bandgap to range from 3.3 eV [9] to 4.6 eV [10]. This variation highlights the sensitivity of the bandgap to the methodological approaches employed and underscores the need for further investigations, particularly under reduced dimensionality. A reliable estimation of their electronic bandgap values is fundamental, as this parameter determines the photon energy required for electron-hole pair separation, a key factor for optoelectronic and catalytic applications [4].

Although vivianites have been extensively investigated in their bulk form, their corresponding 2D structures, vivianenes, remain largely unexplored in the literature [5]. The 2D exfoliated structures enable access to intrinsic properties that facilitate innovative applications in advanced technologies, such as pesticide sensing [5]. However, fundamental aspects of the electronic and optical properties of vivianenes remain to be further investigated.

In this work, we present comprehensive theoretical analyses of vivianenes, focusing on their structural stability, electronic properties, and optical behavior. Using DFT-based *ab initio* molecular dynamics (AIMD) simulations, we compare the properties of vivianites (bulk) and vivianenes (2D), addressing the effects of reduced dimensionality on these characteristics. The objectives are to provide a better understanding of this material and stimulate further experimental and computational research on 2D phosphate materials and their potential applications in energy, sensing, and optoelectronics.

# Computational Details

To investigate the properties of bulk vivianite and its monolayer (010 plane, referred to here as vivianene), we carried out *ab initio* simulations using the Spanish Initiative for Electronic Simulations with Thousands of Atoms (SIESTA) code [11,12]. SIESTA is based on the DFT formalism [13]. We have used the Perdew-Burke-Ernzerhof (PBE) [14] exchange-correlation functional within the Generalized Gradient Approximation (GGA), along with a polarized double-$\zeta$ (DZP) basis set composed of numerical orbitals. A mesh cutoff of 300 Ry was adopted, and the Brillouin zone was sampled using a Γ-centered Monkhorst-Pack grid with 3 × 3 × 5 and 3 × 1 × 5 **k**-points sampling [15] for the bulk and monolayer structures, respectively. For Vivianene, a vacuum buffer region of 25 Å was set along the *y*-direction to prevent spurious interactions with mirror images. Here, the 2D system is oriented along the *xz* plane.

Regarding the self-consistency iteration convergence, we have considered it converged when the difference between the input and output elements of the density matrix was smaller than $10^{-4}$ and when the residual forces were below 0.05 eV/Å. Initially, all calculations were performed at 0 K.

We have also carried out AIMD simulations to assess Vivianene's structural thermal stability and investigated the dynamics of its intrinsic water-like structures. These AIMD runs were performed for approximately 3.5 ps with a time step of 1 fs in an NVT ensemble at 300 K, employing a Nosé-Hoover thermostat [11,12]. The simulations were performed using a mesh cutoff of 250 Ry, and the Brillouin zone was sampled using a 2×1×2 Monkhorst-Pack **k**-point grid [15].

For the optical properties analyses, we have calculated the absorption coefficient (α), reflectivity (R), and refractive index (η) as functions of the photon energy frequency (ω) using the method described in [17] and the following equations:

$$\alpha(\omega) = \sqrt{2}\omega \cdot \left[\left(\epsilon_1^2(\omega) + \epsilon_2^2(\omega)\right)^{1/2} - \epsilon_1(\omega)\right]^{1/2},$$

$$R(\omega) = \left[\frac{(\epsilon_1(\omega) + i\epsilon_2(\omega))^{1/2} - 1}{(\epsilon_1(\omega) + i\epsilon_2(\omega))^{1/2} + 1}\right]^2, \text{ and}$$

$$\eta(\omega) = \frac{1}{\sqrt{2}}\left[\left(\epsilon_1^2(\omega) + \epsilon_2^2(\omega)\right)^{1/2} + \epsilon_1(\omega)\right]^2.$$

**Fig. 1 (a)-(c)** presents the bulk form of vivianite**.** The results of its structural characteristics are summarized in **Table 1**, where DFT-based calculations yielded lattice parameters of a = 9.91 Å, b = 12.87 Å, c = 4.59 Å, along with an angle $\gamma$ = 105.0°. Additionally, the electronic bandgap was calculated to be approximately 3.2 eV. These values agree with previous experimental findings [5], with deviations below 4% for all discussed characteristics, indicating that the computational model employed here provides a reliable description of the vivianite crystalline structure. Furthermore, the results are in good agreement with recent crystallographic data for synthetic vivianite, which reported a = 10.101 Å, b = 13.452 Å, c = 4.711 Å, and $\gamma$ = 104.30° [4].

| System | Vivianite (Exp.) | Vivianite (Theo.) | Vivianene (Theo.) |
|---|---|---|---|
| a | 10.09 [5] | 9.91 | 9.98 |
| b | 13.44 [5] | 12.87 | - |
| c | 4.73 [5] | 4.59 | 4.58 |
| γ | 104.3 [4] | 105.0 | 104.5 |
| Eg | - | 3.21 | 3.03 |
| Egopt | 3.5 [4] | 3.2 | 3.5-3.7 |

## Results and Discussions

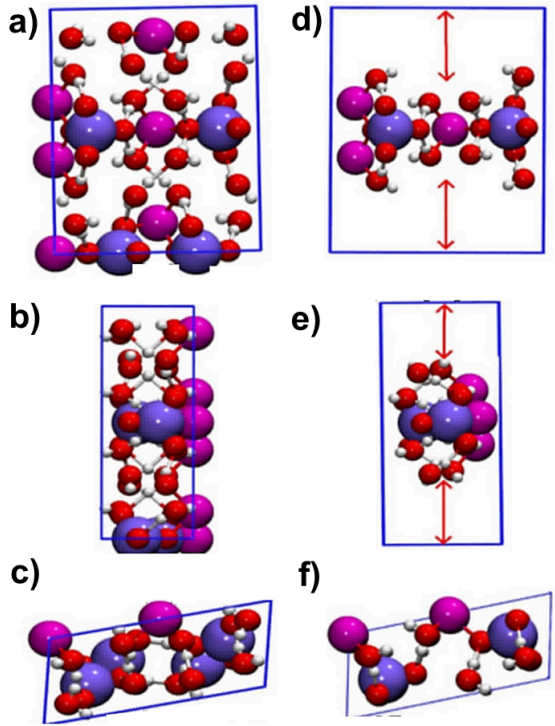

**Figure 1**: Optimized structures of vivianite and its corresponding two-dimensional (2D) form, vivianene. Different crystallographic planes are presented: (a)-(c) for vivianite and (d)-(f) for vivianene. Panels (a), (b), and (c) show the structures along the xy, yz, and xz planes, respectively, for vivianite, while panels (d), (e), and (f) display the corresponding planes for vivianene. In (d) and (e), red arrows indicate the interlayer spacing of the exfoliated structure. Pink, purple, red, and white spheres represent phosphorus (P), iron (Fe), oxygen (O), and hydrogen (H) atoms, respectively.

Vivianene, the monolayer structure shown in **Fig. 1 (d)-(f),** was experimentally obtained by cleaving bulk vivianite along the (010) direction [5]. In this configuration, the central Fe atom interacts with four water-like molecules, and the unit cell is composed of 37 atoms, corresponding to half the number of atoms in the bulk Vivianite unit cell. In the 2D form, the optimized lattice parameters are a = 9.98 Å, c = 4.59 Å, and $\gamma$ = 104.5°. These values are comparable to those obtained for bulk vivianite, with deviations below 1%, indicating that vivianene retains the main structural features of its bulk form. Vivianene consists mainly of [PO$_4$] and [FeO$_6$] clusters, with average P–O bond lengths of 1.57 Å and Fe–O bond lengths of 2.03 Å. Additionally, the free water molecules are, on average, located 3.64 Å from the [PO$_4$] clusters, and the mean O–H bond length is approximately 1.0 Å, consistent with values previously reported in the literature [4].

To further assess the thermodynamic stability of vivianene and analyze the behavior of the water-like structures, we performed AIMD simulations. The results revealed no significant variation in the total energy over 3.5 ps at 300 K, indicating that the system maintains its structural integrity under these conditions. Furthermore, the water-like molecules did not exhibit significant displacement relative to the monolayer. The initial distance between the water molecule and the phosphorus atom in vivianene was 3.64 Å, fluctuating within the range of 3.40–4.15 Å around the original position throughout the simulation time.

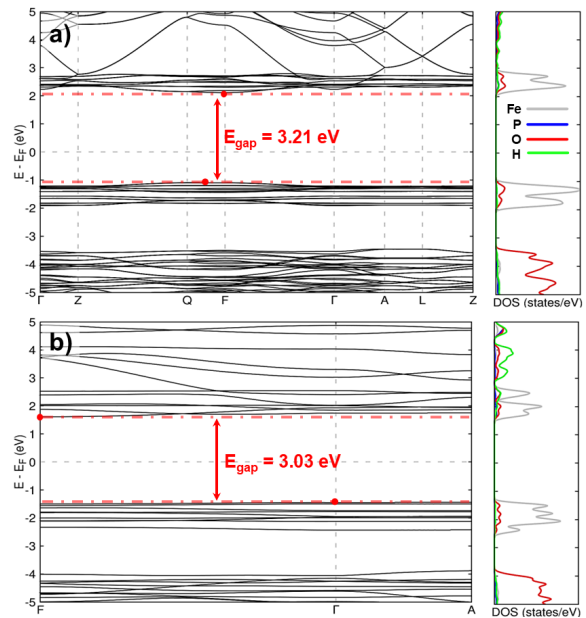

**Figure 2:** Electronic band structure and projected density of states of (a) Vivianite and (b) Vivianene.

Regarding the electronic properties of vivianite and vivianene, we have calculated the electronic band structure and the corresponding projected density of states (PDOS). The results are presented in **Fig. 2.** Both structures exhibit indirect electronic band gaps, with values of 3.21 eV for the bulk structure and 3.03 eV for the monolayer. These results are not typical for the transition from 3D to 2D materials. Generally, quantum confinement effects tend to increase the band gap in 2D structures due to reduced dimensionality. However, similar behavior has also been reported for other materials [10].

**Figure 3:** The absorption coefficient α, refractive index η, and reflectivity R as a function of the photon energy value.

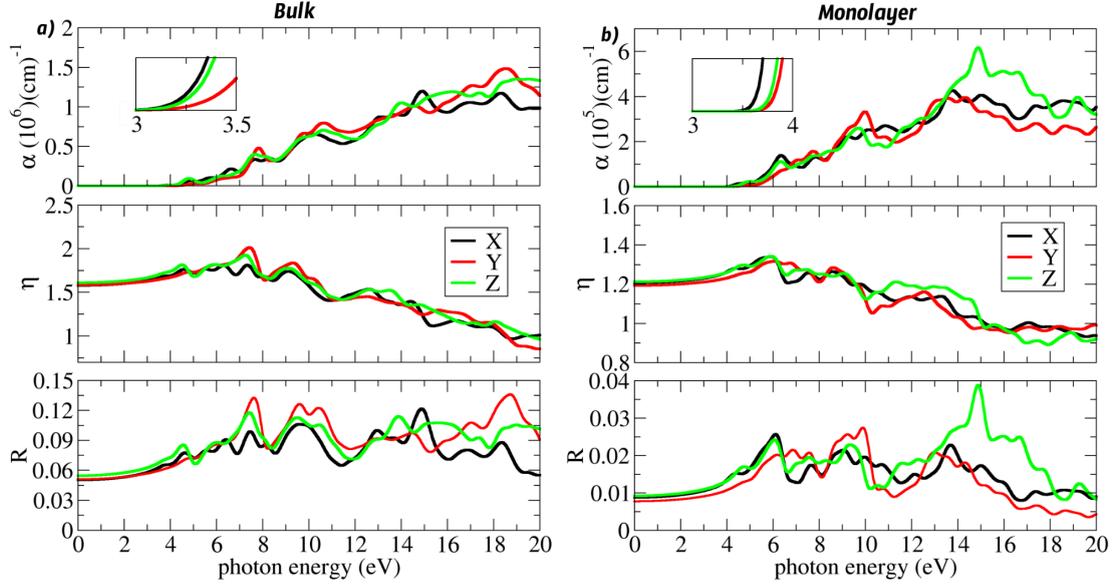

Furthermore, **Fig. 2** shows that the 2D structure exhibits flatter bands compared to the bulk structure, suggesting a reduction in electron mobility. The PDOS displays a similar pattern for both the bulk and monolayer forms. The top and bottom of the valence and conduction bands are primarily composed of d orbitals from Fe atoms, with a smaller contribution from oxygen atoms. The contribution of Fe to the valence bands decreases at higher negative energy values, while the participation of oxygen becomes more significant.

Finally, we discuss the optical properties of the systems investigated in this study. We present the absorption coefficient (α), the refractive index (η), and the reflectivity (R) of Vivianite (**Fig. 3(a)**) and Vivianene (**Fig. 3(b)**) as a function of photon energy, considering the range from 0 to 20 eV. First, it is observed that the absorption coefficient for both the bulk and monolayer structures is nearly isotropic from the onset of absorption up to a photon energy of 13 eV. Although the curves for the three directions in this range are distinct, this does not indicate that absorption is significantly different for any specific light polarization direction. The difference becomes noticeable only for photon energy values above 13 eV, corresponding to the long ultraviolet spectral range.

Regarding absorption intensity, the bulk structure exhibits an absorption intensity that is one order of magnitude larger than that of the monolayer. From an experimental perspective, the absorption intensity depends on how the sample is prepared. However, assuming that Vivianite and Vivianene samples are prepared under equivalent conditions, absorption is significantly higher for the bulk structure across all light polarization directions.

There are no absorption peaks in the infrared region, and absorption in the visible region was negligible, indicating that Vivianite is primarily optically active in the ultraviolet range. In each figure, we also present insets with a magnified area where the systems begin to absorb, with values associated with optical gap transitions. The estimated optical gap values are 3.2 eV for Vivianite, in close agreement with experimental results [4], and 3.6 eV for Vivianene. In this context, quantum confinement effects contribute to the increase in the optical gap of the monolayer compared to the bulk structure.

Regarding the refractive index, the maximum value is approximately 7 eV for Vivianite and 5 eV for Vivianene, tending to decrease as photon energy increases. The maximum intensity of the refractive index decreases from 2.0 eV (1.4 eV) to 1.0 eV (1.0 eV) for the bulk (monolayer) structure. Moreover, the refractive index remains at least one order of magnitude higher than the reflectivity values, even for its minimum value of around 1.0. The maximum reflectivity, on the other hand, is approximately 0.15 for Vivianite and 0.04 for Vivianene. Comparing these values with η, it can be concluded that most of the incident light on the material is absorbed in both the bulk and monolayer forms.

## Conclusions

In this work, we carried out a comprehensive theoretical investigation of Vivianene, the 2D counterpart of bulk Vivianite, focusing on its structural, electronic, and optical properties. Our results show that Vivianene retains the main structural characteristics of its bulk precursor, with minimal deviations in lattice parameters and high thermal stability at room temperature, as confirmed by AIMD simulations.

Electronic structure calculations revealed that both materials exhibit indirect band gaps, with values of 3.21 eV for Vivianite and 3.03 eV for Vivianene. This indicates a deviation from the typical quantum confinement trend in 2D materials. The PDOS analysis showed that Fe d orbitals dominate the valence and conduction bands. The optical response of Vivianene revealed an increased optical band gap (3.6 eV) compared to bulk Vivianite (3.2 eV), with absorption primarily in the ultraviolet region. The refractive index and reflectivity analysis suggest intense optical activity, making Vivianene a promising material for optoelectronic applications.

## Acknowledgement

R.B.O. and B.I. thank CNPq process numbers #151043/2024-8 and #153733/2024-1, respectively. B.I. and G. S. L. F. thank FAPESP process numbers #2024/11016-0 and #2024/03413-9, respectively. M.L.P.J. acknowledges financial support from FAPDF (grant 00193-00001807/2023-16), CNPq (grant 444921/2024-9), and CAPES (grant 88887.005164/2024-00). D. S. G. acknowledges the Center for Computing in Engineering and Sciences at Unicamp for financial support through the FAPESP/CEPID Grant #2013/08293-7.
We thank the Coaraci Supercomputer for computer time (Fapesp grant #2019/17874-0) and the Center for Computing in Engineering and Sciences at Unicamp (Fapesp grant #2013/08293-4).